\documentclass[review]{elsarticle}

\usepackage{lineno,hyperref}
\modulolinenumbers[5]
\usepackage{amsfonts,amscd,amssymb,bm}
\usepackage{amsthm,amsmath,natbib}
\usepackage{bbm} 
\usepackage{times,graphicx}
\usepackage{mathrsfs}
\usepackage{enumerate}
\usepackage{subfigure}
\usepackage{appendix}
\usepackage{pdfpages}
\usepackage[margin=1in]{geometry}
\usepackage{setspace}
\usepackage{tikz}
\usepackage{graphicx}
\usepackage{esvect}
\usepackage{natbib}
\usepackage{algorithm,algcompatible,algpseudocode}
\usepackage{lastpage}
\usepackage{fancyhdr}
\usepackage{color}
\usepackage{pifont}
\usepackage{threeparttable}
\usepackage{tikz}
\usepackage{multirow}

\journal{Journal of \LaTeX\ Templates}

\newcommand{\mc}{\mathcal}

\DeclareMathOperator{\E}{\mathbb{E}}

\newcommand{\tree}{\mathbb{T}}

\newcommand{\var}{\mathrm{Var}}

\newcommand{\tc}[1]{\multicolumn{1}{c}{#1}} 

\begin{document}

\begin{frontmatter}

\title{Statistical modeling for adaptive trait evolution in randomly evolving environment}
\author{Dwueng-Chwuan Jhwueng}
\address{Department of Statistics, Feng-Chia University \\No 100, Seatwen, Taichung Taiwan}

\ead[url]{dcjhwueng@fcu.edu.tw}

\begin{abstract}
In past decades, Gaussian processes has been widely applied in studying trait evolution using phylogenetic comparative analysis.
In particular, two members of Gaussian processes: Brownian motion and Ornstein-Uhlenbeck process, have been frequently used to describe continuous trait evolution.
Under the assumption of adaptive evolution, several models have been created around Ornstein-Uhlenbeck process where the optimum $\theta^y_t$ of a single trait $y_t$ is influenced with predictor $x_t$.
Since in general the dynamics of rate of evolution $\tau^y_t$ of trait could adopt a pertinent process,
in this work we extend models of adaptive evolution by considering the rate of evolution $\tau_t^y$ following the Cox-Ingersoll-Ross (CIR) process.
We provide a heuristic Monte Carlo simulation scheme to simulate trait along the phylogeny as a structure of dependence among species.
We add a framework to incorporate multiple regression with interaction between optimum of the trait and its potential predictors.
Since the likelihood function for our models are intractable, we propose the use of Approximate Bayesian Computation (ABC) for parameter estimation and inference.
Simulation as well as empirical study using the proposed models are also performed and carried out to validate our models and for practical applications.
\end{abstract}


\end{frontmatter}


\section{Introduction}
In statistical phylogenetics, studying how species evolved helps people to understand evolution better.
As many questions are arising from evolutionary biology and ecology, one interesting research question could be: how could traits of a group of related species behave to adapt the changing environment?
For example, when studying marine species\citep{Watanabe12052015}, a scientist may be interested in understanding the moving speed and moving style by comparing fin structures in various kind of swordfish.
One useful tool to track down their evolutionary information is incorporating a phylogenetic tree into analysis.
A phylogenetic tree $\tree$ is a branching diagram that infers evolutionary relationships among a group of species.
Given a tree $\tree$ and traits (e.g. fin lengths or total lengths of fish in center-meter), we could use statistical approach to study ancestral status for species as well as how one trait could be related to the other trait.
From mathematical perspective, changing of trait value or status during evolutionary history can be viewed as a stochastic random variable defined on time/status domain.
In the case of continuous trait, let $y_t$ be a trait of a species observed at time $t$.
The dynamic behavior of $y_t$, when applied for studying trait evolution, can be assumed as a solution to the following stochastic differential equation (SDE)
\begin{equation}\label{eq:ysde}
dy_t= \mu(y_t,\theta,t)dt + \tau(y_t,\theta,t)dW_t,~t>0.
\end{equation}
In the left hand side of Eq. \ref{eq:ysde}, $dy_t$ represents the amount of change in an infinitesimal time $dt$.
In the right hand side of Eq. \ref{eq:ysde}, the deterministic term $\mu(y_t,\theta,t)$ is referred to a drift coefficient that measures the amount of change in an infinitesimal time $dt$ while $\tau(y_t,\theta,t)$ is called the diffusion coefficient that amplifies the trait change according to the random changing environment measured by $dW_t$ where $W_t$ is a Wiener process having continuous paths and independent Gaussian increments (i.e. $dW_t \sim \mc{N}(0, dt)$) and $\theta$ is the model parameters.

In literature, there have been statistical methods developed for traits evolution by applying continuous stochastic processes ranging from Gaussian process (\cite{felsenstein1985, hansen1996}) or non-Gaussian processes \citep{Blomberg067363, jhwueng2019}.
Currently one of the most popular continuous process for trait evolution can be credited to the Ornstein Uhlenbeck(OU) process \citep{hansen1997}.
An OU stochastic variable $y_t$ solves the SDE in Eq. (\ref{eq:ysde}) with $\mu(y_t,\theta,t)=\alpha(\theta-y_t)$ and $\tau(y_t,\theta,t)=\tau$.
The OU process provides a suitable interpretation in describing natural selection in evolution and ecology context.
The constant parameter $\theta$ is intepreted as the optimum status (evolutionary niche in ecology context) of $y_t$.
The parameter $\alpha$ is called a constraining force that pulls trait $y_t$ back to the optimum $\theta$.
The parameter $\tau$ is called the rate of evolution and measures the speed of the random change.

Many works have been developed by expanding the OU model through considering more sophisticated and complex biological phenomenon.
Those models used a generalized OU process to describe trait change along the tree.
The generalized OU model for trait evolution is built by assuming pertinent processes for model parameters $\alpha^y_t, \theta^y_t$ and $\tau^y_t$.
Therefore, the trait $y_t$ solves the following SDE:
\begin{equation}\label{eq:ousde}
dy_t= \alpha^t_y(\theta^y_t-y_t)dt + \tau^y_tdW^y_t,~t>0.
\end{equation}
Currently several works have been focus on the conditions by assuming $\alpha^t_y=\alpha_y$ as a constant, $\theta^y_t$ or $\tau_y^t$ as either a constant or with a stochastic dynamics during the evolutionary process (see \cite{butler2004}, \cite{omeara2006}, and \cite{beaulieu2012}).
By assuming $\theta_t^y$ following a pertinent process, $\theta_t^y$ solves the following SDE:
\begin{equation}\label{eq:thetasde}
d\theta^y_t= \mu(\theta^y_t,t)dt + \sigma(\theta^y_t,t)dW^\theta_t,~t>0.
\end{equation}
In particular, in the case of $\mu(\theta^y_t,t)=0$ and $\sigma(\theta^y_t,t)=\sigma_\theta$, \cite{hansen08} created an OUBM model for optimal regression analysis built under the assumption that the optimum $\theta_y^t$ has a linear relationship with predictors.
\cite{Jhwueng2014} expanded the OUBM model to OUOU model by allowing an Ornstein-Uhlenbeck process for the dynamic of $\theta_y^t$
(i.e. $\mu(\theta^y_t,t)=-\alpha_y(\theta^y_t - \tilde{\theta} )$ and $\sigma(\theta^y_t,t)=\sigma_\theta$).
Those models are applied to study the adaptive relationship of traits building upon its optimum with
$\theta^y_t = \beta_0+\sum_{i=1}^k \beta_i x_{i,t}$
where $\{x_{i,t}\}_{i=1}^k$ is a set of predictors, $\beta_i, i=0,1,\cdots,k$ are regression parameters.
See application sections in \citep{hansen08,Jhwueng2014,jhwueng2016adaptive}.

For the rate evolution $\tau_y^t$ in Eq. (\ref{eq:ousde}), instead of considering constant value or piecewise constant value \citep{omeara2006}, it is also reasonable to assume that the rate of evolution $\tau^y_t$ follows another pertinent process.
Under this assumption, $\tau^y_t$ is a solution to another SDE:
$d\tau^y_t =  \mu(\tau^y_t,\theta,t)dt  + \sigma(\tau^y_t,\theta,t)dW^\tau_t.$
In literature, \cite{jhwueng2016adaptive} considered the rate $\tau_t^y$ to be a Brownian motion where $\mu(\tau^y_t,\theta,t)=0$ and $\sigma(\tau^y_t,\theta,t) = \sigma_\tau$ is a constant.


In this work, observing that there are needs and possibilities to create models for more sophisticated and realistic biological applications, we expand previous existed models in two folds.
First, as the rate $\tau_t^y$ is regarded as non-negative for $t>0$, 
we intend to incorporate a Cox-Ingersoll-Ross(CIR) process \citep{cox1985} for $\tau_t^y$.
In this case, $\tau_t^y$ solves the following SDE:
\begin{equation}\label{eq:cirsde}
d\tau^y_t = \alpha_\tau(\tilde{\tau} - \tau^y_t)dt + \sigma_\tau\sqrt{\tau^y_t}dW^\tau_t
\end{equation}
where $\alpha_\tau>0$ is a constant force, $\tilde{\tau}>0$ is the optimum of $\tau_t^y$ and $\sigma_\tau>0$ is the rate of change for $\tau_t^y$.
In CIR process, the distribution of future values of $\tau_t^y$ conditioned in current value $\tau_s^y$  has a distribution of
$c\chi^2(k,\lambda)$ where $c=\sigma_\tau^2(1-e^{-\alpha_\tau t})/(4\alpha_\tau), k= 4\tilde{\tau}\alpha_\tau/\sigma^2_\tau$
and $\lambda=4\tau_s^y\alpha_\tau e^{-\alpha_\tau t} /
(\sigma^2_\tau(1-e^{-\alpha_\tau t}))$ and $\chi^2(k,\lambda)$ is a non central chi-squared distribution.
Notice that in Eq. (\ref{eq:cirsde}), the diffusion coefficient involves a term $\sqrt{\tau_t^y}$ which indicates that the Eq. (\ref{eq:cirsde}) is neither a linear SDE nor $\tau_t^y$ a normal distributed stochastic variable.
Hence statistical inference on the parameter estimation under our new model will be different from the framework in \citep{hansen08,Jhwueng2014} using the multivariate normal distribution for jointly modeling trait evolution.
Secondly, we assume that there exists an interaction relationship between the optimum $\theta_t^y$ and predictors $x_{i,t}, i=1,2,\cdots,k$ as following
\begin{equation}\label{eq:optint}
\theta^y_t = \beta_0+ \sum_{i=1}^k\beta_i x_{i,t} + \sum_{i,j=1}^k \beta_{ij}x_{i,t}x_{j,t}
\end{equation}
where the term $x_{i,t}x_{j,t}$ is the interaction between the $i$th and the $j$th predictors with regression parameter $\beta_{ij}$.
Note that this model is different from the phylogenetic ancova model in \cite{fuen16} where the optimum $\theta_t^y$ is not considered with relationship to the predictors as shown in Eq. (\ref{eq:optint}).

When jointly modeling adaptive trait evolution using Eqs. (\ref{eq:ousde}), (\ref{eq:thetasde}), (\ref{eq:cirsde}) and (\ref{eq:optint}), the distribution for the trait $y_t$ of a species is constrained by the dynamic assumption of the rate parameter $\tau_t$ via either a Brownian motion case \citep{jhwueng2016adaptive} or the CIR process case in Eq. (\ref{eq:cirsde}).

However, $y_t$ given $\tau_t^y$ as a CIR process is not a Gaussian random variable and has intractable model likelihood.
Conceiving this, we propose an algorithm under the approximation Bayesian computation(ABC) framework for statistical inference.
We describe our framework into the following sections.
Section \ref{sec:model} illustrates the general construction of adaptive model under a various of assumption of pertinent processes for $y_t, \theta^y_t,$ and $\tau^y_t$.
We call our new models the OUBMCIR model for $y_t$ following generalized OU process with $\theta^y_t$ following a BM and $\tau_t^y$ following a CIR process.
And we called the OUOUCIR model for $y_t$ following a generalized OU process with $\theta^y_t$ following an OU process and $\tau_t^y$ following a CIR process.
Section \ref{sec:simtrait} contains methods on simulating traits under each model.
We make an attempt to derive the solution $y_t$ as explicitly as possible for the purpose of applying tree traversal algorithm \citep{fel2004} to simulate trait status on the internal nodes and tips on the tree.
We conduct statistical inference for parameter estimation under ABC in Section \ref{sec:abcalg}.
Currently we mainly use the \texttt{R} package \texttt{abc} for inference after traits are simulated from Section \ref{sec:sims}.
We provide empricial analsis on analyzing data from literature in Section \ref{sec:emp}.
We conclude our study in Section \ref{sec:cls}.
The scripts and their brief description developed in work project can be accessed at Github: \url{https://github.com/djhwueng/ououcir}.

\section{Model}\label{sec:model}
\subsection{Property of adaptive trait models}\label{sec:nomassum}
We start this section by first introducing some definitions of the SDE property.
In Eq. (\ref{eq:ysde}), the SDE is a linear SDE if $\mu(y_t,t)= a_1(t)y_t + a_2(t)$ and $\tau(y_t,t)= b_1(t)y_t+b_2(t)$ are linear function of $y_t$.
That is,
$dy_t=(a_1(t)y_t+a_2(t))dt + (b_1(t)y_t + b_2(t))dW_t.$
A linear SDE is autonomous if all coefficients are constants, is homogenous if $a_2(t)=0$ and $b_2(t)=0$ and is linear in the additive sense if $b_1(t) = 0$.

These properties could provide some information on the distribution of $y_t$.
For instance, the SDE for $y_t$ in OUBM model \citep{hansen08} with $\mu(y_t,t)=\alpha(\theta_t-y_t)$ and $\tau_t^t = \tau $ is a linear additive non-autonomous SDE.
In the OUBM model, since both $\theta_t$ and $W_t$ are BMs, the solution for the SDE in Eq. (\ref{eq:ysde}) is represented as a linear combination of two BMs.
As dynamics of each BM can be treated as a normal random variable, we can conclude that $y_t$ is normal random variable in OUBM model.
In this case, we can implement normal distribution to analyze data.
We categorize the properties of SDE of $y_t$ as well as $\theta_t$ and $\tau_t$ in Table \ref{tb:prosde}.
  \begin{table}[ht]\centering
  \begin{tabular}{rlllll}
    \hline
   parameters& $(y_t,\theta_t,\tau_t)$ & $(y_t,\theta_t,\tau_t)$ & $(y_t,\theta_t,\tau_t)$ & $(y_t,\theta_t,\tau_t)$ &  \\
   Model  & Linear & Autonomous & Additive & Normal & References \\
   \hline
    OUBM & (\checkmark, \checkmark, -) & (n, \checkmark, -) & (\checkmark, \checkmark, -) & (\checkmark, \checkmark, -) & \citep{hansen08} \\
    OUOU & (\checkmark, \checkmark, -) & (n, \checkmark, -) & (\checkmark, \checkmark, -) & (\checkmark, \checkmark, -) & \citep{Jhwueng2014} \\
    OUBMBM & (\checkmark, \checkmark, \checkmark) & (n, \checkmark, \checkmark) & (\checkmark, \checkmark, \checkmark) & (n, \checkmark, \checkmark) & \citep{jhwueng2016adaptive} \\
    OUOUBM & (\checkmark, \checkmark, \checkmark) & (n, \checkmark, \checkmark) & (\checkmark, \checkmark, \checkmark) & (n, \checkmark, \checkmark) & \citep{jhwueng2016adaptive} \\
    OUBMCIR & (\checkmark, \checkmark, n) & (n, \checkmark, n) & (\checkmark, \checkmark, n) & (n, \checkmark, n) & This work \\
    OUOUCIR & (\checkmark, \checkmark, n) & (n, \checkmark, n) & (\checkmark, \checkmark, n) & (n, \checkmark, n) & This work \\
     \hline
  \end{tabular}
  \caption{Property of adaptive trait models.
  The check symbol \checkmark represents a yes for the property, and the letter n represents a no and the symbol - means not available.
  The term $(\cdot,\cdot,\cdot)$ refers to the property of SDE for the triple parameters $(y_t,\theta_t,\tau_t)$.
  For instance, in the OUBMBM model the triple parameters $(y_t,\theta_t,\tau_t)$ with (\checkmark,\checkmark,\checkmark) in linearity property (Linear) has a meaning that all of them are solution to a linear SDE.
  On the other hand, the SDE for $y_t$ in OUOUCIR model with (\checkmark,\checkmark,n) is a linear non-autonomous, additive SDE where the solution $y_t$ is not a normal distributed stochastic variable.
  }
  \label{tb:prosde}
  \end{table}

\subsection{Solution of Model }\label{sec:solmodel}
In general, by adopting Eqs. (\ref{eq:ousde}), (\ref{eq:thetasde}), (\ref{eq:cirsde}) and (\ref{eq:optint}), we can present the dynamic of $y_t,\theta_t^y, \tau_t^y$ into a system of SDE for the random vector $\bm{Z}_t=(y_t,\theta_t^y,\tau_t^y)^t$ as
$d\bm{Z}_t= \bm{\mu}_tdt + \bm{D}_td\bm{W}_t$,
where $\bm{\mu}_t = (\mu(y_t,t), \mu(\theta^y_t,t), \mu(\tau_t,t))^T$ is the drift vector, $\bm{D}_t = \text{diag}\left[\tau(y_t,t),\sigma(\theta^y_t,t),\sigma(\tau_t^y,t)\right]$ is the diffusion vector, and $\bm{W}_t=(W_t^y,W^{\theta}_t,W^{\tau}_t)^T$ is the associated independent Brownian process random vector and $v^T$ is a transpose of a vector $v$.
By assuming that the force parameters are time invariant($\alpha_t=\alpha$), the model can be represented as
\begin{equation} \label{eq:linsde}
d\bm{Z}_t= (\bm{A}\bm{Z}_t + \bm{b}_t)dt + \bm{D}_td\bm{W}_t.
\end{equation}

For \textit{homogeneous} model assuming the rate of evolution $\tau^y_t$ as a time invariant constant (i.e. $\bm{b}_t=0$ and $\tau_y^t=\tau_y$ in OUBM model and in OUOU model), we have $\alpha_\tau=0$ and $\bm{D}_t= \text{diag}\left[ \tau, \sigma_\theta, 0\right]$ is a constant diagonal matrix.
In this case, given the initital condition $\bm{Z}_0=(y_0,\theta_0,\tau_0^y)^t$ at
$t=0$, the system of SDE described by Eq. (\ref{eq:linsde}) has a unique solution $\bm{Z}_t=  e^{-\bm{A}t} \bm{Z}_0 + \int_0^t e^{-\bm{A}(t-s)}\bm{D}_s d\bm{W}_s$.
In this case, the expected value of $\bm{Z}_t$ can be calculated straightforwardly as
$\E[\bm{Z}_t]= \bm{Z}_0 e^{-\bm{A}t}$ while the second moment of the random vector $\bm{Z}_t$, denoted by $\bm{P}_t=\E[\bm{Z}_t\bm{Z}_t^T]$, can uniquely be determined by solving the system of an ordinary differential equation
$\frac{d}{dt}\bm{P}_t =\bm{AP}_t+\bm{P}_t\bm{A}^T+ \mathbb{E}[\bm{C}_t]$
where $\E[\bm{C}_t] = \bm{S}_t\bm{S}_t$.
Once the first and second moment of $\bm{Z}_t$ are identified, because $\bm{Z}_t$ is a normal random vector, its first component $y_t$ is a normal random variable.
We can also work from Eq. (\ref{eq:ousde}) on the assumption that $\tau_t=\tau$ is a constant.
The solution $y_t=y_0 e^{-\alpha_y t} + \alpha_y e^{-\alpha_y t}\int_0^t e^{\alpha_y s} \theta_s ds + \tau \int_0^t e^{- \alpha_y (t-s)} dW_s^y$ is a linear combination of normal random variable which is again a normal random variable under the assumption of BM for $\theta_t$ \citep{hansen08} or OU for $\theta_t$ \citep{Jhwueng2014}.

On the other hand, however, for OUBMBM, OUOUBM, OUBMCIR, and OUOUCIR model, as the rate of evolution $\tau_t^y$ follows a certain pertinent process, the distribution of $\bm{Z}_t$ is not as straightforward to work through.
We show that $\bm{Z}_t$ fails to be a normal distributed random vector.
We first demonstrate this using the new proposed OUOUCIR model with
\begin{equation}\notag
  \bm{D}_t= \left( \begin{array}{ccc}
  \tau^y_t&0&0\\
  0&\sigma_\theta&0\\
  0&0&\sigma_\tau \sqrt{\tau_t}\\ \end{array} \right),
\end{equation}
and
\begin{equation}\notag 
   \bm{\mu}_t=\bm{AZ}_t+\bm{b}_t= \left( \begin{array}{ccc}
   -\alpha_y&\alpha_y&0\\
   0&-\alpha_\theta&0\\
   0&0&-\alpha_\tau\\ \end{array} \right)\bm{Z}_t+
  \begin{pmatrix}
  0
  \\
  0
  \\
  \alpha_\tau \tilde{\tau}
  \end{pmatrix}.
\end{equation}
Due to assumption of using CIR process for the rate parameter $\tau_t^y$ and the stationary distribution of a CIR random variable is not a normal random variable,
the solution to the system of equation in Eq. (\ref{eq:linsde}) is intractible and not likely to be normal distribution.

Moreover, even for $\tau_t$ following a Brownian motion, we claims that $y_t$ fails to be a normal random variable.
For the OUBMBM model in \citep{jhwueng2016adaptive}, the solution $y_t$ for the SDE in Eq. (\ref{eq:ousde}) under OUBMBM model is
\begin{equation}\label{eq:oubmbm}
y_t = y_0 + e^{-\alpha_y t} \int_0^t \alpha_y e^{\alpha_y s} \theta_s ds +
e^{\alpha_y t} \int_0^t \tau_s e^{\alpha_y s} dW_s^y = y_0 + \textcircled{1} + \textcircled{2}
\end{equation}
where $\theta_s=\sigma_\theta W_s^\theta$ and $\tau_s=\sigma_\tau W_s^\tau$ are standard Wiener processes.

By direct calculation on the stochastic integral, we have $\textcircled{1} = \sigma_\theta \int_0^t [e^{\alpha_y t} - \alpha_y e^{\alpha_y s}]dW_s^\theta$ which is a normal random variable with mean $0$ and variance (by It\^{o}'s isometry) $\sigma^2_\theta(t e^{2\alpha_y t} - 2[e^{2\alpha_y t} -e^{\alpha_y t}] +
\frac{\alpha_y}{2}[e^{2\alpha_y t} -1])$.
However, the stochastic integral in \textcircled{2} fails to be a normal random variable.
To show this, for simplicity we assume that $W_s^\tau$ and $W_s^y$ are two identical and independent Wiener processes.
Let $f(s,W)=e^{\alpha_y (s-t)} W^2$, by It\^{o}'s lemma we have
$\textcircled{2} = \sigma_\tau \frac{1}{2} e^{\alpha_y t} W_t^2 - \sigma_\tau \frac{\alpha_y}{2}\int_0^t e^{\alpha_y s} W_s^2 ds + \sigma_\tau \frac{e^{\alpha_y t} -1}{2\alpha_y}$.
Since neither $W_t^2$ nor $\int_0^te^{\alpha_y s}W_s^2ds$ is a normal random variable, \textcircled{2} fails to be a normal distributed.
This indicates that to $y_t$ in Eq. (\ref{eq:oubmbm}) can not a normal random variable.

\subsection{Multiple optimal regression with interaction}
In this section, we describe how to implement the interaction in Eq. (\ref{eq:optint}) into our model.
To start, we use an example of two predictors $x_{1,t},x_{2,t}$ for illustration.
The general case can be extended accordingly.
Given that the linear relationship between the optimum $\theta_y^t$ and predictors with interaction is $\theta^y_t= b_0 + b_1 x_{1,t} + b_2 x_{2,t} + b_{12}x_{1,t}x_{2,t}$, by differentiating on both side of the equation with respect to $t$, we have
\begin{equation}\label{eq:dtheta}
d\theta_t^y = b_1dx_{1,t}+b_2dx_{2,t} + b_{12}[x_{2,t}dx_{1,t} + x_{1,t}dx_{2,t} + dx_{1,t} dx_{2,t}]
\end{equation}
where $x_{i,t}$ is a diffusion process satisfies the SDE as following
\begin{equation}\label{eq:xsde}
dx_t= \mu(x_t,t)dt + \sigma(x_t,t)dW^x_t,~t>0.
\end{equation}
By the SDE of $\theta_y^t$ in Eq. (\ref{eq:thetasde}) and assumptions of stochastic calculus with $dtdt \approx 0, dtdW_t \approx 0, dW_tdW_t \approx dt$, we have
$d\theta^y_t d \theta^y_t = \sigma^2(\theta^y_t,t)dt$.
In the case of assuming $\theta_t^y$ either a BM or an OU process, we have $\sigma(\theta_t^y,t)=\sigma_\theta$ which implies $d\theta^y_t d\theta^y_t  = \sigma^2_\theta dt$.
Similarly for $x_t$ in Eq. (\ref{eq:xsde}) for either BM or OU process, we have $\sigma(x_t,t)=\sigma_x$ and $(dx_t)^2=\sigma_x^2dt$.
The relationship between $\sigma_\theta$ and $\sigma_x$ given the predictor traits $x_{1,t}$ and $x_{2,t}$ can be derived  with expanding $d\theta_t^y d \theta_t^y$ using Eq. (\ref{eq:dtheta}) and represented as
\[\sigma^2_\theta = \sigma_{x_{1}}^2 ( b_1^2 +2b_1b_{12}x_{2,t}  +b_{12}^2x_{2,t}^2) +   \sigma_{x_{2}}^2 (b_2^2 +2b_2b_{12}x_{1,t}  + b_{12}^2x_{1,t}^2).\]
The general case of optimum regression on the predictors with interaction can be extended from above with assumption with the form
\begin{equation}\label{eq:gentheta}
\theta_t^y = b_0 + \sum_{k=1}^n b_kx_{k,t} + \sum_{i\neq j} b_{ij}x_{i,t}x_{j,t}.
\end{equation}
By applying the same technique from above, we have
\begin{equation}\label{eq:genDtheta}
d\theta_t^y = \displaystyle\sum_{k = 1}^n b_k\sigma_{x_k}dW_{t}^{x_k} + \displaystyle\sum_{i = 1}^{n}\sum_{i \neq j}^{n}b_{ij}(x_{j,t}\sigma_{x_{i}}dW_t^{x_i} + x_{i,t}\sigma_{x_j}dW_t^{x_j} + \rho_{ij}\sigma_i\sigma_jx_{i,t}x_{j,t}dt)
\end{equation}
where $-1 \leq \rho_{ij}\leq 1$ is the correlation between two Wiener processes (i.e. $dW_t^{x_i}dW_t^{x_j}= \rho_{ij} dt$).

Then using the same technique on $d\theta_t^y d\theta_t^y$ and compare it with $dx_{i,t}dx_{i,t}$, we have
\begin{equation}\label{eq:sigmatheta}
\sigma^2_\theta = \displaystyle\sum_{i = 1}^{n}b_i^2\sigma^2_{x_i} + \sum_{i = 1}^{n}\sigma_{x_i}^2\sum_{j\neq i}^n b_{ij}^2x_{j,t}^2 + 2\sum_{i = 1}^{n}b_i\sigma_{x_i}^2\sum_{j \neq i}^n b_{ij}x_{j,t}.
\end{equation}

Eq. (\ref{eq:sigmatheta}) suggests that $\sigma^2_\theta$ depends on the predictors $x_{i,t}$s which are stochastic variable, in order to quantify $\sigma_\theta^2$, we consider to use expected value of $\sigma^2_\theta$.

When $x_t$ is a Brownian motion, since $\E[x_t]=0$ and $\E[x_t^2]=\sigma_x^2 t$, we have

\begin{equation}\label{eq:exbm}
\E[\sigma^2_\theta] = \displaystyle\sum_{i = 1}^{n}b_i^2\sigma^2_{x_i} + \sum_{i = 1}^{n}\sigma_{x_i}^2\sum_{j\neq i}^n b_{ij}^2\sigma_{x_i}^2 t.
\end{equation}

When $x_t$ is an OU process, we have
\begin{equation}\label{eq:exou}
\E[\sigma^2_\theta] =\displaystyle\sum_{i = 1}^{n}b_i^2\sigma^2_{x_i} + \sum_{i = 1}^{n}\sigma_{x_i}^2\sum_{j\neq i}^n b_{ij}^2\E[x_j^2] + 2\sum_{i = 1}^{n}b_i\sigma_{x_i}^2\sum_{j \neq i}^n b_{ij}\E[x_j]
\end{equation}
where $\E[x_t]=x_0 \exp(-\alpha_x t) + \mu_x(1-\exp(-\alpha_x t))$ and $\E[x_t^2] = \sigma^2[1-\exp(-2 \alpha_x t)]/(2\alpha_x) + [x_0 \exp(-\alpha_x t)+ \mu(1-\exp(-\alpha_x t))]^2$.

\section{Simulate trait along tree}\label{sec:simtrait}
Given a tree $\tree$ with known topology and length, we simulate tip as well as ancestral states using tree traversal algorithm \citep{fel2004} under a specified model $\mc{M}$.
In particular, when the distribution is known, for instance, under Brownian motion trait value of a species at time $t$ conditioned on its ancestor $y_a$ on $\tree$ is a normal random variable $y_t|y_a$ with mean $y_a$ and variance $\sigma^2
t$.  (i.e. $y_t|y_a \sim \mc{N}\left(y_a,\sigma^2t\right)$).
Under OU process, $y_t|y_a$ is a normal random variable with mean $y_0 e^{-\alpha_t} + \theta (1- e^{-\alpha t})$, and variance $\sigma^2(1-e^{-2\alpha_t})/(2\alpha))$.
Moreover, under either BM or OU process the tip can be simulated directly under the joint distribution (i.e.
$\bm{Y} \sim \mc{N}(\bm{\mu}, \sigma^2\bm{\Sigma}_\alpha)$ where $\bm{Y}=(y_1,y_2,\cdots,y_n)^n$ is the trait vector at tip of the tree, $\bm{\mu}$ is the mean vector, and $\bm{\Sigma}_\alpha$ is the variance covariance structure for $\bm{Y}$\citep{jhwueng13}).

Given the prior information on model parameters, our goal is to simulate $i$th response trait $y_i, i=1,2,\cdots,n$, and predictor traits $x_{i,m}, m=1,2,\cdots,m$ at the tip.
We describe our method for simulating trait under each model using the given parameters values.

\subsection{OUBM \& OUOU model}
For OUBM model, the model parameters are $\alpha_y, \sigma_y,$ and $\sigma_x$, and regression parameters are $b_i, b_{ij}, i,j=1,2,\cdots, n$.
We first simulate predictor traits $x_i$s on each node/tip of tree given $\sigma_x$.
The optimal value $\theta_i$ can then be calculated via $\theta_i = \sum b_ix_i + \sum b_{ij}x_ix_j$ given $b_i$ and $b_{ij}$.
Then use $\alpha_y, \sigma_y$ to simulate $y|y_a \sim \mc{N}(\E[y|y_a], \var(y|y_a))$ (see \citep{hansen08} for the formula of $\E[y|y_a]$ and $\var(y|y_a)$).

For OUOU model, model parameters are $\alpha_y, \sigma_y, \alpha_x, \theta_x$, and $\sigma_x$, and regression parameters are $b_is, b_{ij}s, i,j = 1,2,\cdots,n.$
We simulate predictor trait $x_i$s on each node/tip of tree using $\alpha_x,\theta_x,\sigma_x$.
The optimal value can be calculated via $\theta_i = \sum b_ix_i + \sum b_{ij}x_ix_j$ to obtain $\theta$ on each nodes.
We use $\alpha_y,\sigma_y$ to simulate $y_t$, by $y_t|y_a \sim \mc{N}(\E[y|y_a], \var(y|y_a))$ (see \citep{Jhwueng2014} for the formula of $\E[y|y_a]$ and $\var(y|y_a)$).

Note that since the OUBM model and OUOU model are both of multivariate normal distributions, trait values at tips $\bm{Y}$ can be simulated directly given the specified mean vector $\E[\bm{Y}]$ and variance structure $\var[\bm{Y}]$.

\subsection{OUBMBM model}
In OUBMBM model, the model parameters are $\alpha_y, \tau, \sigma_x$, and regression parameters are $b_i, b_{ij}, i,j=1,2,\cdots,n$.
We first simulate predictor traits $x_i$s on each node/tip of tree given $\sigma_x$.
The optimal value $\theta_i$ can then be calculated via $\theta_i = \sum b_ix_i + \sum b_{ij}x_ix_j$ given $b_i$ and $b_{ij}$.
To simulate $y_i$s at the nodes/tips, we first look at the solution in Eq. (\ref{eq:ousde}) for $y_t$:
\begin{equation}\label{eq:solytoubmbm}
y_t= y_0 + e^{-\alpha_y t} \int_0^t \alpha_y e^{\alpha_y s} \theta_s ds + e^{-\alpha_y t}\int_0^t \tau_s e^{\alpha_y s} dW_s^y
= y_0 + \textcircled{1} + \textcircled{2}.
\end{equation}

For \textcircled{1}, as we assume the optimum follows Brownian motion (i.e. $\theta_s=\int_0^s \sigma_\theta dW_v^\theta = \sigma_\theta W^\theta_s \sim \mc{N}(0,\sigma_\theta^2 s)$), the term
$\int_0^t \alpha_y e^{\alpha_y s} \theta_s ds$ is a stochastic integral of Brownian motion with respect to time and equals to $\int_0^t \alpha_y e^{\alpha_y s} \theta_s ds = \int \theta_s d e^{\alpha_y s}$.

Since $d (\theta_s e^{\alpha_y s}) = e^{\alpha_y s} d\theta_s + \theta_sde^{\alpha_y s}$, we have
the integral  $\int \theta_s d e^{\alpha_y s}  =\int_0^t d(\theta_s e^{\alpha_y s}) - \int_0^t e^{\alpha_y s} d\theta_s =
\theta_t e^{\alpha_y t} - \theta_0 - \int_0^t e^{\alpha_y s} d\theta_s$ which is a normal random variable with mean $\theta_t e^{\alpha_y t} - \theta_0$ and variance $\frac{e^{2\alpha_y}-1}{2 \alpha_y}$

In \textcircled{2}, since the rate is assumed as BM (i.e. $\tau_s=\int_0^s \sigma_\tau dW_v^\tau = \sigma_\tau W^\tau_s \sim \mc{N}(0,\sigma_\tau^2 s)$), we have $\textcircled{2} = e^{-\alpha_y t}\int_0^t \tau_s e^{\alpha_y s} dW_s^y =\int_0^t \sigma_\tau W_s^\tau e^{\alpha_y(s-t)}dW_s^y$.
Hence \textcircled{2} is a stochastic integral that involves an
integral of Brownian motion $W_s^\tau$ with respect to another Brownian motion
$W_s^y$.
Note \textcircled{2} is not a normal distributed random variable (see section \ref{sec:nomassum}).
In order to draw sample from \textcircled{2}, we use function \texttt{int.st} in \texttt{R} package \texttt{Sim.DiffProc} \citep{guid17} to simulate the trajectory of this stochastic integral.
We assume $W_s^y$ and $W_s^\sigma$ are two independent and identical processes.
We then use median of the trajectory as a sample for \textcircled{2}.
Given the parameter values, we can apply tree traversal algorithm to simulate sample $y_i$ on node/tip conditioned on its ancestor $y_a$.

\subsection{OUOUBM model}
In OUOUBM model, model parameters are $\alpha_y, \alpha_x, \theta_x, \sigma_x, \tau$, and regression parameters $b_i, b_{ij}, i,j=1,2,\cdots,n$.
We first simulate predictor traits $x_i$s on each node/tip of tree using $\alpha_x,\theta_x,\sigma_x$.
The optimum on each node and tip can be calculated as $\theta_i = \sum b_ix_i + \sum b_{ij}x_ix_j$.
To simulate $y_i$s, since the solution in Eq. (\ref{eq:ousde}) under OUOUBM model is
\begin{equation}\label{eq:solytououbm}
y_t= y_0 + e^{-\alpha_y t} \int_0^t \alpha_y e^{\alpha_y s} \theta_s ds + e^{-\alpha_y t} \int_0^t \sigma_s e^{\alpha_y s} dW_s^y = y_0 + \textcircled{1} + \textcircled{2}.
\end{equation}

For \textcircled{1}, because $\theta_s$ is an OU process with $\theta_s=e^{-\alpha_\theta s}\theta_0 + \theta_1(1-e^{-\alpha_\theta s}) +\sigma_\theta \int_0^s  e^{\alpha_\theta (v-s)}dW^\theta_v$ where $\theta_1$ is optimum of $\theta_s$ and $\theta_0$ is the initital condition.
The integral $\int_0^t \alpha_y e^{\alpha_y s} \theta_sds$ becomes
\begin{equation}\label{eq:solthetaou}
\int_0^t \alpha_y \theta_0 e^{(\alpha_y-\alpha_\theta)s} ds
+ \int_0^t \alpha_y\theta_1e^{\alpha_y s}(1-e^{-\alpha_\theta s})   ds
+ \int_0^t \sigma_\theta \alpha_y e^{(\alpha_y-\alpha_\theta )s} \left(\int_0^s e^{\alpha_\theta v}d W_v^\theta\right)  ds
= \textcircled{a} + \textcircled{b} + \textcircled{c}.
\end{equation}
Note that \textcircled{a} and \textcircled{b} are both definite integrals with \textcircled{a} = $\frac{\alpha_y \theta_0}{\alpha_y-\alpha_\theta}(e^{(\alpha_y - \alpha_\theta)t}-1)$ and \textcircled{b} = $\theta_1 (e^{\alpha_y t} -1) - \frac{\alpha_y \theta_1}{\alpha_y-\alpha_\theta}(e^{(\alpha_y-\alpha_\theta)t}-1)$.
In \textcircled{c},
the term $\int_0^s e^{\alpha_\theta v} dW_v^\theta$ is a normal random variable with mean $0$ and variance $\frac{e^{2\alpha_\theta s} -1 }{2\alpha_\theta}$.
The integrand in \textcircled{c} defined as $f_s=\sigma_\theta \alpha_y e^{(\alpha_y-\alpha_\theta )s} \left(\int_0^s e^{\alpha_\theta v}d W_v^\theta\right)$ is a normal random variable with mean $0$ and variance (by It\^{o} Isometry) $ v(s) =
\frac{\sigma_\theta^2\alpha_y^2}{2\alpha_\theta}(e^{2\alpha_y s} - e^{2(\alpha_y-\alpha_\theta)s})$.
So \textcircled{c} = $\int_0^t f_{v(s)}ds$ is again a normal random variable.
Because $v(s)$ is not an invertible function, it is not likely to identify the distribution of $\int_0^t f_{v(s)}ds$ directly using change of variable.
We alternatively use linear approximation for $v(s)$ with
$v(s)=a+bs$ at $s=0$ where $a=q(0)=0$ and $b=q'(0)=\sigma_\theta^2 \alpha_y^2 s$ to obtain an candidate of distribution of $\int_0^t f_{v(s)}ds \approx \int_0^t
f_{\sigma_{\theta^2 \alpha_y^2s}}ds$ which is a normal random variable with mean $0$ and variance $(\sigma_\theta^2\alpha_y^2 t)^3/(3\sigma_\theta^2\alpha_y^2)$.

For \textcircled{2}, as the rate is a BM, we can simulate samples use the method for the \textcircled{2} described in the OUBMBM model.

\subsection{OUBMCIR model}
In OUBMCIR model, the model parameters are $\alpha_y, \sigma_x, \alpha_\tau, \tilde{\tau}, \sigma_\tau$, and regression parameters are $b_is$, $b_{ij}s,
i,j=1,2,\cdots,n$.
We first use $\sigma_x$ to simulate predictor trait $x_i$s and then use
$\theta_i = \sum b_ix_i + \sum b_{ij}x_ix_j$ to obtain $\theta_i$ on each node/tip.
To simulate $y_i$s, since the solution in Eq. (\ref{eq:ousde}) is

\begin{equation}\label{eq:solytoubmcir}
y_t= y_0 + e^{-\alpha_y t} \int_0^t \alpha_y e^{\alpha_y s} \theta_s ds + e^{-\alpha_y t}  \int_0^t \tau_s e^{\alpha_y s} dW_s^y = y_0 + \textcircled{1} +
\textcircled{2}.
\end{equation}

For \textcircled{1}, since the optimum is a BM (i.e $\theta_s \sim \mc{N}(0,
\sigma^2_\theta s)$), we can draw using the expected value and variance as shown in the \textcircled{1} in the OUBMBM model.

For \textcircled{2}, it is a stochastic integral of a CIR random variable $\tau_s$ with respect to Brownian motion $W_s^y$.
Note that $\tau_s|\tau_0$ follows a scaled non-central chi-squared distribution
$c \chi^2 (k,\lambda)$ where $c = \sigma_\tau^2(1 - e^{-\alpha_\tau t}) / (4 \alpha),
k = 4 \tilde{\tau} \alpha_\tau/\sigma_\tau^2, \lambda = 4\tau_0 \alpha_\tau
e^{-\alpha_\tau t}/(\sigma_\tau^2(1 - e^{-\alpha_\tau t})$ and $\chi^2_{k,\lambda}$ is a non-central chi-squared distribution with degree of freedom $k$ and non-centrality parameter $\lambda$ \citep{jhwueng2019}.

The distribution of the random variable $\int_0^t \tau_s e^{\alpha s}dW_s^y$ conditioned on $\tau_0$ can be seen as the sum of three independent random variables (see \textit{prop}. 4 Eq. 2.10 in \cite{chan10}).
Moreover, \cite{glasserman11} and \cite{chan10} showed that the exact distribution of $\int_0^t \tau_sds$, conditional on $\tau_t$ and $\sigma_0$ can be representd by infinite sums and mixtures of gamma random variables (see prop 4. in \citep{chan10}). For our case, to simulate sample in \textcircled{2}, we first simulate
$\tau_s$ on each node along the tree using tree traversal as in \citep{jhwueng2019}.
We next simulate sample for the random variable $\int_0^t\tau_s e^{\alpha_y s} dW_s^y$.
Since the solution to the CIR SDE in Eq. (\ref{eq:cirsde}) is given by
\begin{equation}\label{eq:solcirtau}
\tau_s= \tilde{\tau} + (\tau_0-\tilde{\tau})e^{-\alpha_\tau s} + \sigma_\tau e^{-\alpha_\tau s} \int_0^s e^{\alpha_\tau u} \sqrt{\tau_u}dW_u.
\end{equation}
The integral $\int_0^t\tau_s e^{\alpha_y s} dW_s^y $ can be separated into three parts: \textcircled{a} + \textcircled{b} +\textcircled{c}.
For $\textcircled{a} = \int_0^t \tilde{\tau} e^{\alpha_y s} dW_s^y$, it is a normal random variables with mean 0 and variance $\tilde{\tau}^2 \frac{e^{2\alpha_y t}-1}{2\alpha_y}$.
For $\textcircled{b} = (\tau_0 - \tilde{\tau})\int_0^t e^{(\alpha_y -\alpha_\tau)s} dW_s^y$, it is another normal random variable with mean 0 and variance $(\tau_0-\tilde{\tau})^2 (e^{2(\alpha_y-\alpha_\tau)t}-1)/(2(\alpha_y-\alpha_\tau)) $.
For \textcircled{c} = $\sigma_\tau \int_0^t e^{(\alpha_y-\alpha_\tau) s} \left(\int_0^s e^{\alpha_\tau u}\sqrt{\sigma_u} dW_u^\sigma \right) dW_s^y$, unfortunately, it has no analytical distribution.
We instead try to use numerical approach to draw sample.
To illustrate this, let $x_s=  \int_0^s e^{\alpha_\tau u} \sqrt{\tau_u} dW_u^\tau$. We use \texttt{st.int} function in \citep{guid17} to calculate this stochastic integral on the interval $[0,s]$ where $\sigma_{u_i}$ on the subintervals $(s_i,s_{i+1})$ is a noncentral chi-square random variable.
Then we simulate samples $\tau_{u_i}$ on the subinterval $[s_i,s_{i+1}]$ and draw sample $W_i$ from normal distribution with mean 0, and variance $s_{i+1}-s_i$. Then $x_{s,j}$ is sampled by the sum $\sum_{i=0}^n{_j} e^{\alpha_\tau s_i} \sqrt{\sigma_{u_i}}W_i$.
Eventually we obtain a sample for \textcircled{c} using the sum
$\sum_{j=1}^m e^{-\alpha_\tau s_j}  x_{s,j}v_j $ where $v_j$ is a normal random variable with mean 0 and variance $t_{i+1} - t_i$.

\subsection{OUOUCIR model}
In OUOUCIR model, the model parameters are $\alpha_y, \alpha_x, \theta_x,\sigma_x, \alpha_\tau, \tilde{\tau}, \sigma_\tau$, and regression parameters $b_i, b_{ij}, i,j =1,2,\cdots,n.$. We first use $\alpha_x,\theta_x,\sigma_x$ to simulate predictor trait $x_i$s and then use $\theta_i = \sum b_ix_i + \sum b_{ij}x_ix_j, i,j = 1,2,\cdots,n$ to obtain $\theta_i$ on each node/tip.
To simulate $y_i$s, since the solution for $y_t$ in Eq. (\ref{eq:ousde}) for OUOUCIR model is
\begin{equation}\label{eq:solytououcir}
y_t= y_0 + e^{-\alpha t} \int_0^t \alpha e^{\alpha s} \theta_s ds + e^{-\alpha t} \int_0^t \tau_s e^{\alpha s} dW_s^y = y_0 + \textcircled{1} + \textcircled{2}.
\end{equation}
We can use the same method for the \textcircled{1} described in OUOUBM model to simulate the sample for \textcircled{1} and use the same method for the \textcircled{2} described in OUBMCIR model to simulate samle for \textcircled{2}.

Note that \cite{goolsby2017} developed a two-pass algorithm to perform ancestral reconstruction and applied to multivariate trait evolution, non-Brownian models, missing data and phylogenetic regression. In the near future, we could develop possible more efficient algorithm for drawing samples.

\section{Inference} \label{sec:abcalg}
\subsection{Approximate Bayesian Computation for adaptive trait model}
As mentioned in section \ref{sec:nomassum}, we cannot specifiy the distribution of $y_t$ for OUBMBM, OUOUBM, OUBMCIR and OUOUCIR models.
To do statisdtical inference on the parameters of interest, we propose to use
Approximate Bayesian Computation (ABC) approach.
Our goal is to compute the posterior probability distribution for the model parameters, says, $\Theta$.
To start ABC approach, a parameter vector $\Theta_i$ is drawn under its joint prior distribution.
We first simulate replicates of trait $\bm{Y}_i, i=1,2,\cdots,m$ under model $\mc{M}$.
Then a set of summary statistics $S(\bm{Y}_i)$ are computed from the simulated data and compared with the summary statistics of the raw data $S(\bm{Y})$ using a distance measure $d$.
In general, $d$ is the Euclidean distance between two summary statistics.
Note that before computing the distance, \cite{Blum2010} suggests to scaled each summary statistics by a robust estimate of the standard deviation (the median absolute deviation).
If the distance between $S(\bm{Y}_i)$ and $S(\bm{Y}_0)$ is less than a given threshold $\delta$ (i.e. $d(S(\bm{Y}_i),S(\bm{Y}))<\delta$), then the drawed parameter vector $\Theta_i$ is accepted.

In fact, we need to establish a procedure for choosing good summary statistics for ABC.
ABC fails to be accurate when using too many summary statistics as the distance increases with the number of summary statistics.
The inference could be more accurate with high efficiency if we use the summary statistics that utilizes the all data info.
To attain this goal, we would focus on choosing summary statistic on a pragmatic basis by making use of tree $\tree$ and trait $\bm{Y}$ so that the statistics could capture the important model's behavior.
In phylogenetic comparative analysis, we might want to capture the overall amount of evolution, the over-dispersion of trait values, and the phylogenetic structuring of the trait values.
\cite{clark17} used the mean and the variance of the differences between each species and its closet neighbor in trait space for BM and OU model as the summary statistics.
As our model falls out of the exponential family of distributions, it is theoretical impossible to quantify all finite dimensional sufficient statistics.
However, it still possible to implement non-sufficient statistics when inference is under the ABC framework.

Currently, we consider to use the \textbf{mean} and the \textbf{variance of the differences between each species} suggested in \cite{clark17}.
We will continue to look for more possible sufficient summary stastistic so our inference will be more efficient with reduced error.
After choosing appropriate summary statistics, a tolerance rate defined as the percentage of accepted simulation is provided for the aids to set up the threshold value.
Then the posterior distribution of the parameters can be approximated using the accepted $\Theta_i$s.
Furthermore, \cite{Blum2010a} implemented a regression adjustment to improve the estimation of posterior distribution via weaken the effect of the discrepancy between the observed summary statistics and the accepted ones.
The aims for this additional step is to rectify the match between the accepted summary statistics $S(\bm{Y}_i)$ and observed summary statistics $S(\bm{Y})$.
The regression equation for the adjustment can be written as $\theta_i = m(S(\bm{Y}_i)) + \epsilon_i$ where $m$ is a regression function, and $\epsilon_i$s are centered random variables with a common variance.
Once the regression is performed, a weighted sample from the posterior distribution is obtained by correcting the $\theta_i$s via
$\theta_i^*=\hat{m}(S(\bm{Y}))+\hat{\epsilon}_i$, where $\hat{m}(\cdot)$ is the estimated conditional mean and the $\hat{\epsilon}_i$s are the empirical residuals of the regression \citep{Beaumont2025}.
Additionally, a correction for heteroscedasticity is applied
$\theta_i^*= \hat{m}(S(\bm{Y})) +  (\hat{\sigma}(S(\bm{Y}))/\hat{\sigma}(S(\bm{Y}_i))) \hat{\epsilon}_i$ where $\hat{\sigma}(\cdot)$ is the estimated conditional standard deviation \citep{Blum2010a}.
We provide a more detail description of our modeling procedure using ABC algorithm in Algorithm \ref{alg:abc}.
\begin{algorithm}[H]
  \caption{Approximate Bayesian Computation rejection method for OUBMBM, OUOUBM, OUBMCIR and OUOUCIR models.}
  \label{alg:abc}
  \begin{algorithmic}[1]
    \Require
      Tree $\mathbb{T}$ with branch length and topology, initial state $\theta_0$, trait data $Y, X_1, X_2$, prior distribution $\pi(\theta)$, a tolerance $\epsilon$.
    \Ensure
      Posterior sample $\theta_i$, $i = 1, 2, \dots, k$ from posterior distribution.
     \For{$i = 1, \dots, k$}
      \State simulate sample $\theta_i$ from $\pi(\theta_0)$.
      \State simulate trait $Y_i,X_{1i},X_{2i}$ form $\theta_i$.
      \State compute the distance $d_i$ between two summary statistics $S(Y_i)$ and $S(Y)$
      \If{$d_i < \epsilon$}
      \State accept $\theta_i$;
      \Else
      \State reject $\theta_i$.
      \EndIf
     \EndFor \\
    \Return $\theta_i$, $i = 1, 2, \dots, k$.
  \end{algorithmic}
\end{algorithm}

\subsection{Model selection under ABC} \label{subsec:abcms}
Currently, for the posterior samples under rejection method, we use the function \texttt{postpr} in \texttt{abc} package \citep{csillery2012abc} to computes the posterior model probabilities where the posterior probability of a given model is approximated by the proportion of accepted simulations given this model.
This approximation holds when the different models are a prior equally likely, and the same number of simulations is performed for each model.
We then compute the Bayes factors (BF) to compare a pair of models in the model sets.
From conventional statistics on the definition of the Bayes factor which is a ratio of the likelihood probability of two competing hypotheses, usually a null and an alternative.
The posterior probability $\text{Pr}(M|D)$ of a model $M$ given data $D$ is given by Bayes' theorem:
\[\text{Pr}(M|D) = \frac{\text{Pr}(D|M)\text{Pr}(M)}{\text{Pr}(D)}.\]
Given a model selection we have to choose between two models on the basis of observed data $D$, the plausibility of the two different models $M_1$ and $M_2$, parametrised by model parameter vectors $\theta_1$ and $\theta_2$ is assessed by the Bayes factor $K$ given by
\[K=\frac{\text{Pr}(D|M_1)}{\text{Pr}(D|M_2)} = \frac{\int\text{Pr}(\theta_1|M_1)\text{Pr}(D|\theta_1,M_1)\text{d}\theta_1}{\int\text{Pr}(\theta_2|M_2)\text{Pr}(D|\theta_2,M_2)\text{d}\theta_2} = \frac{\text{Pr}(M_1|D)}{\text{Pr}(M_2|D)}
\frac{\text{Pr}(M_2)}{\text{Pr}(M_1)}.\]
A value of $K > 1$ means that $M_1$ is more strongly supported by the data than $M_2$.
For models where an explicit version of likelihood is not available or too costly to evaluate numerically, approximate Bayesian computation can be used for model selection in a Bayesian framework, with the caveat that approximate-Bayesian estimates of Bayes factors are often biased.
Here as we use ABC and we do not have likelihood function.
We read the \texttt{R} script \texttt{postpr} function \citep{csillery2012abc} which interprets the algorithm to compute the Bayes factor like a version for model selection.
For our works, we have 4 models where each model contains 50,000 replicates data.
We first compute the Euclidean distance for each replicate with respect to the realization(true data).
By setting the acceptance rate, we decide the cutoff of the distance calculated by the scaled summary statistics.
We then grasp and count the frequency of each model that has the distance smaller than this cutoff.
Eventually, the Bayes factor between two models is computed as the ratio using the frequencies of two models.

For instance, with the acceptance rate of 10 percent. We will expect 5000 replicates among the 50000*4=200000 replicated for all model.
We sort the 200000 distance and determine the cutoff at the 5000th position.
We then count the frequency of each model that has the distance smaller than the cutoff.
For example, OUBMBM has 1200, OUOUBM has 1500 OUBMCIR has 1800, and OUOUCIR has 500.
Then the Bayes factor of OUOUBM with respect to OUOUCIR is 3.
\cite{kass1995} suggested that a value $K$ more than 150 would show very strong support for model 1 over model 2, between 20 and 150 would show strong support for model 1 over model 2, between 3 and 20 show positive support for model 1 over model 2, finally a value $K$ between 1 and 3 could not worth more than a bare mention for model1 and model 2.

\section{Simulation}\label{sec:sims}
We consider using different informative prior for simulation, and different sampling approach.
We have four models (OUBMBM, OUOUBM, OUBMCIR, and OUOUCIR) where every model has different parameters for itself.
For simulation, we set the true parameters for the four model as following $\alpha_y = 0.15, \alpha_x=0.1, \theta_x=0, \sigma_x=1, \alpha_\tau=0.2, \theta_\tau=30, \tau=0.35, \sigma_\tau=0.5, b_0 = 0, b_1 = 0.5, b_2 = 0.5$.
We set the prior distribution parameters are $\alpha_y\sim U(0,0.3), \alpha_x\sim U(0,0.2), \theta_x\sim U(-1,1),
\sigma_x\sim U(0,2), \alpha_\tau \sim U(0,0.4), \theta_\tau \sim U(0,60), \sigma_\tau \sim U(0,1), \tau\sim U(0,0.7), b_0\sim U(-1,1), b_1\sim U(0,1), b_2\sim U(0,1)$.
We run fifty thousand replicates in the simulation that have four models and four taxa size(10, 20, 50 and 100) and generate four different model tables containing bias of parameters estimates, standard deviation, and 90\% confidence interval.
Next, the previous assumptions of prior distribution are the uniform distribution, then we will try to set the different informative prior distribution for simulation.
We set the prior distribution to $\alpha_y \sim \text{exp}(\frac{1}{0.15}), \sigma_x \sim \text{exp}(1), \tau \sim \text{exp}(1), \alpha_x
\sim \text{exp}(\frac{1}{0.1}), \theta_x \sim N(0,1), \alpha_\tau \sim \text{exp}(
\frac{1}{0.2}), \theta_\tau \sim\chi^2_{30}, \sigma_\tau \sim \text{exp}(\frac{1}{0.5}), b_0\sim U(-1,1), b_1\sim U(0,1), b_2\sim U(0,1)$.
We run fifty thousand replicates in this simulation and output our results in our tables.
Finally, we change the sampling approach, so consider the Approximate Bayesian Computation using Markov chain Monte Carlo (ABC-MCMC), assume the prior distribution and true parameters are the same as ABC rejection method.
We run fifty thousand replicates in the simulation, set the threshold $\delta$ is 100 and burn-in time is 5000, because the first steps of the algorithm may be biased by the initial value, and are therefore usually discarded for the analysis.

\subsection{OUBMBM Model}\label{sub:oubmbm}

\begin{table}[H]
\centering
\caption{OUBMBM model: Bias, Standard deviation and 90\% interval for parameters $ \alpha_y, \sigma_x, \tau, b_0, b_1, b_2$ with uniform prior use rejection approach.}
\begin{tabular}{ccrrrrccrrrr}

& $n$ & \tc{10} & \tc{20} & \tc{50} & \tc{100} & & $n$ & \tc{10} & \tc{20} & \tc{50} & \tc{100}\\
Par.& & & & & & Par.& & & & \\ \hline

\multirow{4}{*}{$\alpha_y$}
&  bias & 0.000 & 0.003 & 0.003 & 0.005 & \multirow{4}{*}{$b_0$}& bias & 0.055 & 0.011 & 0.058 & 0.007\\
&  sd & 0.088 & 0.086 & 0.087 & 0.086 & &  sd & 0.579 & 0.575 & 0.579 & 0.576\\
&  5\% & 0.015 & 0.017 & 0.015 & 0.014 & &  5\% & -0.895 & -0.905 & -0.889 & -0.890\\
&  95\% & 0.286 & 0.285 & 0.285 & 0.283 & &  95\% & 0.910 & 0.898 & 0.910 & 0.901\\ \hline
\multirow{4}{*}{$\sigma_x$}
&  bias & 0.108 & 0.134 & 0.044 & 0.177& \multirow{4}{*}{$b_1$} &  bias & 0.012 & 0.110 & 0.081 & 0.027 \\
&  sd & 0.347 & 0.286 & 0.235 & 0.229& &  sd & 0.284 & 0.272 & 0.281 & 0.271 \\
&  5\% & 0.423 & 0.509 & 0.624 & 0.521& & 5\% & 0.053 & 0.034 & 0.068 & 0.053 \\
&  95\% & 1.566 & 1.441 & 1.382 & 1.265& &  95\% & 0.941 & 0.908 & 0.958 & 0.926 \\ \hline
\multirow{4}{*}{$\tau$}
&  bias & 0.001 & 0.012 & 0.003 & 0.000& \multirow{4}{*}{$b_2$} &  bias & 0.058 & 0.071 & 0.083 & 0.011 \\
&  sd & 0.204 & 0.203 & 0.204 & 0.203& & sd & 0.281 & 0.276 & 0.278 & 0.275 \\
&  5\% & 0.036 & 0.028 & 0.032 & 0.036& & 5\% & 0.067 & 0.040 & 0.070 & 0.050 \\
&  95\% & 0.666 & 0.663 & 0.659 & 0.671& & 95\% & 0.955 & 0.918 & 0.960 & 0.931 \\ \hline
\end{tabular}
\label{tb:OUBMBMrej_u}
\end{table}

In table \ref{tb:OUBMBMrej_u}, we have six parameters in the OUBMBM model, the true parameters values $(\alpha_y, \sigma_x, \tau, b_0, b_1, b_2) = (0.15, 1, 0.35, 0, 0.5, 0.5)$.
The model is so complicated, so we can not estimate easily, bias doesn't keep getting smaller when the size becomes larger.
But the standard deviation is kept getting smaller and the 90\% confidence interval is also narrower as the size becomes larger and larger.

\begin{table}[H]
\centering
\caption{OUBMBM model: Bias, Standard deviation and 90\% interval for parameters $ \alpha_y, \sigma_x, \tau, b_0, b_1, b_2$ with non-information prior and use rejection approach.}
\begin{tabular}{ccrrrrccrrrr}

  & $n$ & \tc{10} & \tc{20} & \tc{50} & \tc{100} & & $n$ & \tc{10} & \tc{20} & \tc{50} & \tc{100}\\
  Par.& & & & & & Par.& & & & \\ \hline
  \multirow{4}{*}{$\alpha_y$}
  & bias & 0.048 & 0.043 & 0.056 & 0.058& \multirow{4}{*}{$b_0$}& bias & 0.028 & 0.029 & 0.012 & 0.007 \\
  & sd & 0.145 & 0.151 & 0.143 & 0.136& & sd & 0.572 & 0.575 & 0.572 & 0.577\\
  & 5\% & 0.006 & 0.008 & 0.008 & 0.008& & 5\% & -0.902 & -0.881 & -0.883 & -0.891 \\
  & 95\% & 0.428 & 0.473 & 0.429 & 0.419& & 95\% & 0.884 & 0.881 & 0.909 & 0.904\\ \hline
  \multirow{4}{*}{$\sigma_x$}
  & bias & 0.017 & 0.148 & 0.299 & 0.256& \multirow{4}{*}{$b_1$}& bias & 0.015 & 0.067 & 0.008 & 0.030 \\
  & sd & 0.398 & 0.301 & 0.226 & 0.256& & sd &0.282 & 0.277 & 0.284 & 0.286\\
  & 5\% & 0.512 & 0.483 & 0.401 & 0.396& & 5\% & 0.059 & 0.039 & 0.051 & 0.053 \\
  & 95\% & 1.828 & 1.448 & 1.136 & 1.244& & 95\% & 0.944 & 0.909 & 0.937 & 0.946 \\ \hline
  \multirow{4}{*}{$\tau$}
  & bias & 0.342 & 0.369 & 0.334 & 0.373& \multirow{4}{*}{$b_2$}& bias & 0.012 & 0.022 & 0.005 & 0.077 \\
  & sd & 1.019 & 0.971 & 1.029 & 1.037& & sd & 0.285 & 0.281 & 0.286 & 0.279\\
  & 5\% & 0.054 & 0.052 & 0.044 & 0.054& & 5\% & 0.055 & 0.045 & 0.054 & 0.070 \\
  & 95\% & 3.107 & 2.995 & 2.987 & 3.081& & 95\% & 0.947 & 0.936 & 0.950 & 0.961 \\ \hline
\end{tabular}
\label{tb:OUBMBMrej_nu}
\end{table}
The table \ref{tb:OUBMBMrej_nu} shows the parameters, bias, standard deviation (sd) and 90\% confidence interval.
Only the bias value of $b_0$ keeps getting smaller when the size gets larger.

\subsection{OUOUBM Model}\label{sub:ououbm}
\begin{table}[H]
\caption{OUOUBM model: Bias, Standard deviation and 90\% interval for parameters $\alpha_y, \alpha_x, \theta_x, \sigma_x, \tau, b_0, b_1, b_2$ with uniform prior use rejection approach.}
\begin{tabular}{ccrrrrccrrrr}
& $n$ & \tc{10} & \tc{20} & \tc{50} & \tc{100} & & $n$ & \tc{10} & \tc{20} & \tc{50} & \tc{100}\\
Par.& & & & & & Par.& & & & \\ \hline
\multirow{4}{*}{$\alpha_y$}
 & bias & 0.013 & 0.044 & 0.010 & 0.006& \multirow{4}{*}{$\tau$}
 & bias & 0.004 & 0.007 & 0.002 & 0.006 \\
 & sd & 0.073 & 0.063 & 0.068 & 0.067& & sd & 0.200 & 0.202 & 0.204 & 0.199 \\
 & 5\% & 0.038 & 0.084 & 0.059 & 0.054& & 5\% & 0.035 & 0.031 & 0.040 & 0.040 \\
 & 95\% & 0.269 & 0.287 & 0.279 & 0.274& & 95\% & 0.663 & 0.667 & 0.665 & 0.666 \\ \hline
\multirow{4}{*}{$\alpha_x$}
 & bias & 0.004 & 0.002 & 0.002 & 0.004& \multirow{4}{*}{$b_0$}
 & bias & 0.035 & 0.083 & 0.022 & 0.001 \\
 & sd & 0.057 & 0.058 & 0.057 & 0.057& & sd & 0.576 & 0.578 & 0.579 & 0.581 \\
 & 5\% & 0.013 & 0.010 & 0.010 & 0.012& & 5\% & -0.912 & -0.895 & -0.896 & -0.901 \\
 & 95\% & 0.190 & 0.191 & 0.190 & 0.191& & 95\% & 0.892 & 0.909 & 0.900 & 0.904 \\ \hline
\multirow{4}{*}{$\theta_x$}
 & bias & 0.028 & 0.004 & 0.036 & 0.044& \multirow{4}{*}{$b_1$}
 & bias & 0.055 & 0.074 & 0.000 & 0.024 \\
 & sd & 0.569 & 0.574 & 0.580 & 0.575& & sd & 0.278 & 0.273 & 0.281 & 0.276 \\
 & 5\% & -0.896 & -0.898 & -0.891 & -0.906& & 5\% & 0.048 & 0.082 & 0.047 & 0.051 \\
 & 95\% & 0.888 & 0.898 & 0.912 & 0.887& & 95\% & 0.927 & 0.954 & 0.950 & 0.939 \\ \hline
\multirow{4}{*}{$\sigma_x$}
 & bias & 0.008 & 0.128 & 0.066 & 0.112& \multirow{4}{*}{$b_2$}
 & bias & 0.030 & 0.001 & 0.030 & 0.037 \\
 & sd & 0.397 & 0.420 & 0.317 & 0.316& & sd & 0.281 & 0.280 & 0.280 & 0.277 \\
 & 5\% & 0.473 & 0.347 & 0.519 & 0.462& & 5\% & 0.053 & 0.057 & 0.042 & 0.051 \\
 & 95\% & 1.789 & 1.733 & 1.550 & 1.493& & 95\% & 0.944 & 0.949 & 0.940 & 0.935 \\ \hline
\end{tabular}
\label{tb:OUOUBMrej_u}
\end{table}
For table \ref{tb:OUOUBMrej_u}, the true parameter values ($\alpha_y, \alpha_x, \theta_x, \sigma_x, \tau, b_0, b_1, b_2$) = (0.15, 0.1, 0, 1, 0.35, 0, 0.5, 0.5).
In this table, the $\alpha_y$ and $b_0$ bias is smaller than other sizes when size is 100, this is we expect the result.

\begin{table}[H]
\caption{OUOUBM model: Bias, Standard deviation and 90\% interval for parameters $\alpha_y, \alpha_x, \theta_x, \sigma_x, \tau, b_0, b_1, b_2$ with information prior and use rejection approach.}
\begin{tabular}{ccrrrrccrrrr}
& $n$ & \tc{10} & \tc{20} & \tc{50} & \tc{100} & & $n$ & \tc{10} & \tc{20} & \tc{50} & \tc{100}\\
Par.& & & & & & Par.& & & & \\ \hline
\multirow{4}{*}{$\alpha_y$}
& bias & 0.002 & 0.076 & 0.031 & 0.037& \multirow{4}{*}{$\tau$}
& bias & 0.112 & 0.114 & 0.098 & 0.109 \\
& sd & 0.109 & 0.070 & 0.080 & 0.117& & sd & 0.342 & 0.344 & 0.362 & 0.341 \\
& 5\% & 0.050 & 0.012 & 0.041 & 0.072& & 5\% & 0.016 & 0.019 & 0.016 & 0.020 \\
& 95\% & 0.372 & 0.222 & 0.292 & 0.438& & 95\% & 1.045 & 0.997 & 1.046 & 1.033 \\ \hline
\multirow{4}{*}{$\alpha_x$}
& bias & 0.032 & 0.031 & 0.027 & 0.032& \multirow{4}{*}{$b_0$}
& bias & 0.020 & 0.018 & 0.011 & 0.093 \\
& sd & 0.095 & 0.099 & 0.105 & 0.098& & sd & 0.575 & 0.580 & 0.578 & 0.566 \\
& 5\% & 0.005 & 0.005 & 0.005 & 0.005& & 5\% & -0.906 & -0.907 & -0.894 & -0.909 \\
& 95\% & 0.291 & 0.305 & 0.315 & 0.289& & 95\% & 0.895 & 0.899 & 0.894 & 0.856 \\ \hline
\multirow{4}{*}{$\theta_x$}
& bias & 0.021 & 0.028 & 0.022 & 0.023& \multirow{4}{*}{$b_1$}
& bias & 0.016 & 0.033 & 0.000 & 0.015 \\
& sd & 1.008 & 0.988 & 0.967 & 0.978& & sd & 0.288 & 0.283 & 0.282 & 0.288 \\
& 5\% & -1.633 & -1.664 & -1.585 & -1.608& & 5\% & 0.059 & 0.057 & 0.053 & 0.050 \\
& 95\% & 1.693 & 1.607 & 1.618 & 1.650& & 95\% & 0.956 & 0.947 & 0.947 & 0.953 \\ \hline
\multirow{4}{*}{$\sigma_x$}
& bias & 0.302 & 0.178 & 0.299 & 0.252& \multirow{4}{*}{$b_2$}
& bias & 0.016 & 0.041 & 0.001 & 0.089 \\
& sd & 0.454 & 0.444 & 0.328 & 0.601& & sd & 0.280 & 0.284 & 0.287 & 0.286 \\
& 5\% & 0.256 & 0.369 & 0.333 & 0.163& & 5\% & 0.059 & 0.041 & 0.046 & 0.072 \\
& 95\% & 1.658 & 1.766 & 1.366 & 2.075& & 95\% & 0.951 & 0.935 & 0.952 & 0.963 \\ \hline
\end{tabular}
\label{tb:OUOUBMrej_nu}
\end{table}

The true parameter values ($\alpha_y, \alpha_x, \theta_x, \sigma_x, \tau, b_0, b_1, b_2$) = (0.15, 0.1, 0, 1, 0.35, 0, 0.5, 0.5). The results of table \ref{tb:OUOUBMrej_nu} are not good, because we expect the bias value and interval range keep getting smaller when the size gets larger.
Therefore, there is no significant difference to change the prior distribution information for the OUOUBM model.

\subsection{OUBMCIR Model}\label{sub:oubmcir}
\begin{table}[H]
\caption{OUBMCIR model: Bias, Standard deviation and 90\% interval for parameters $\alpha_y, \sigma_x, \alpha_\tau, \theta_\tau, \sigma_\tau, b_0, b_1, b_2$ with uniform prior use rejection approach.}
\begin{tabular}{ccrrrrccrrrr}
& $n$ & \tc{10} & \tc{20} & \tc{50} & \tc{100} & & $n$ & \tc{10} & \tc{20} & \tc{50} & \tc{100}\\
Par.& & & & & & Par.& & & & \\ \hline
\multirow{4}{*}{$\alpha_y$}
& bias & 0.001 & 0.006 & 0.001 & 0.001& \multirow{4}{*}{$\sigma_\tau$}& bias & 0.007 & 0.005 & 0.019 & 0.009 \\
& sd & 0.085 & 0.086 & 0.086 & 0.087& & sd & 0.292 & 0.286 & 0.289 & 0.290 \\
& 5\% & 0.015 & 0.017 & 0.018 & 0.015& & 5\% & 0.044 & 0.053 & 0.057 & 0.048 \\
& 95\% & 0.282 & 0.285 & 0.286 & 0.284& & 95\% & 0.956 & 0.944 & 0.955 & 0.943 \\ \hline
\multirow{4}{*}{$\sigma_x$}
& bias & 0.128 & 0.169 & 0.080 & 0.182& \multirow{4}{*}{$b_0$}& bias & 0.008 & 0.008 & 0.003 & 0.003 \\
& sd & 0.359 & 0.348 & 0.289 & 0.255& & sd & 0.572 & 0.570 & 0.573 & 0.571 \\
& 5\% & 0.583 & 0.629 & 0.513 & 0.445& & 5\% &-0.891 & -0.890 & -0.901 & -0.897 \\
& 95\% & 1.777 & 1.768 & 1.450 & 1.276& & 95\% & 0.903 & 0.891 & 0.885 & 0.896\\ \hline
\multirow{4}{*}{$\alpha_\tau$}
& bias & 0.007 & 0.009 & 0.013 & 0.014& \multirow{4}{*}{$b_1$}& bias & 0.013 & 0.000 & 0.013 & 0.015 \\
& sd & 0.116 & 0.115 & 0.116 & 0.114& & sd & 0.286 & 0.289 & 0.288 & 0.288 \\
& 5\% & 0.021 & 0.021 & 0.021 & 0.026& & 5\% & 0.053 & 0.051 & 0.057 & 0.052 \\
& 95\% & 0.384 & 0.383 & 0.384 & 0.383& & 95\% & 0.949 & 0.951 & 0.954 & 0.955 \\ \hline
\multirow{4}{*}{$\theta_\tau$}
& bias & 11.087 & 6.625 & 6.489 & 2.499& \multirow{4}{*}{$b_2$}& bias & 0.000 & 0.030 & 0.007 & 0.009 \\
& sd & 14.244 & 15.051 & 11.641 & 12.102& & sd & 0.290 & 0.289 & 0.288 & 0.286 \\
& 5\% & 2.024 & 2.795 & 6.427 & 10.472& & 5\% & 0.047 & 0.052 & 0.051 & 0.057 \\
& 95\% & 48.393 & 51.861 & 45.130 & 49.624& & 95\% & 0.945 & 0.947 & 0.950 & 0.951 \\ \hline
\end{tabular}
\label{tb:OUBMCIRrej_u}
\end{table}
In table \ref{tb:OUBMCIRrej_u}, the $\alpha_y, \theta_\tau, b_0$ of bias result are smaller than other sizes when size is 100.
And $\theta_\tau$ of the OUBMCIR model is the best estimate compared to other parameters, when the size gets bigger and bigger it bias value is keep getting smaller and the 90\% confidence interval is getting narrower, too.

\begin{table}[H]
\caption{OUBMCIR model: Bias, Standard deviation and 90\% interval for parameters $\alpha_y, \sigma_x, \alpha_\tau, \theta_\tau, \sigma_\tau, b_0, b_1, b_2$ with non-information prior and use rejection approach.}
\begin{tabular}{ccrrrrccrrrr}
 & $n$ & \tc{10} & \tc{20} & \tc{50} & \tc{100} & & $n$ & \tc{10} & \tc{20} & \tc{50} & \tc{100}\\
 Par.& & & & & & Par.& & & & \\ \hline
 \multirow{4}{*}{$\alpha_y$}
 & bias & 0.041 & 0.048 & 0.049 & 0.048& \multirow{4}{*}{$\sigma_\tau$}& bias & 0.148 & 0.157 & 0.155 & 0.158 \\
 & sd & 0.151 & 0.150 & 0.153 & 0.157& & sd & 0.507 & 0.486 & 0.493 & 0.440 \\
 & 5\% & 0.009 & 0.007 & 0.007 & 0.007& & 5\% & 0.028 & 0.025 & 0.028 & 0.029 \\
 & 95\% & 0.445 & 0.455 & 0.448 & 0.456& & 95\% & 1.545 & 1.484 & 1.484 & 1.383 \\ \hline
 \multirow{4}{*}{$\sigma_x$}
 & bias & 0.006 & 0.129 & 0.308 & 0.259& \multirow{4}{*}{$b_0$}& bias & 0.003 & 0.009 & 0.028 & 0.023 \\
 & sd & 0.400 & 0.362 & 0.291 & 0.312& & sd & 0.564 & 0.573 & 0.578 & 0.569 \\
 & 5\% & 0.511 & 0.419 & 0.352 & 0.338& & 5\% & -0.884 & -0.904 & -0.899 & -0.896 \\
 & 95\% & 1.797 & 1.577 & 1.293 & 1.356& & 95\% & 0.901 & 0.893 & 0.900 & 0.890 \\ \hline
 \multirow{4}{*}{$\alpha_\tau$}
 & bias & 0.053 & 0.071 & 0.057 & 0.056& \multirow{4}{*}{$b_1$}& bias & 0.008 & 0.006 & 0.014 & 0.012 \\
 & sd & 0.212 & 0.195 & 0.198 & 0.198& & sd & 0.291 & 0.287 & 0.289 & 0.289 \\
 & 5\% & 0.011 & 0.011 & 0.012 & 0.010& & 5\% & 0.044 & 0.050 & 0.046 & 0.051 \\
 & 95\% & 0.662 & 0.568 & 0.578 & 0.585& & 95\% & 0.951 & 0.947 & 0.951 & 0.945 \\ \hline
 \multirow{4}{*}{$\theta_\tau$}
 & bias & 1.585 & 0.095 & 1.006 & 1.597& \multirow{4}{*}{$b_2$}& bias & 0.002 & 0.003 & 0.002 & 0.020 \\
 & sd & 6.614 & 6.531 & 5.089 & 5.191& & sd & 0.289 & 0.287 & 0.288 & 0.292 \\
 & 5\% & 18.773 & 20.347 & 21.582 & 20.490& & 5\% & 0.054 & 0.050 & 0.050 & 0.045 \\
 & 95\% & 40.291 & 41.735 & 38.094 & 37.479& & 95\% & 0.944 & 0.952 & 0.947 & 0.946 \\ \hline
\end{tabular}
\label{tb:OUBMCIRrej_nu}
\end{table}

In table \ref{tb:OUBMCIRrej_nu}, we mainly attention to parameters $\alpha_y, \sigma_x, \alpha_\tau, \theta_\tau, \sigma_\tau$, because the prior distribution information of these parameters is changed. But the OUBMCIR model is complex, so we cannot estimate these parameters easily.
The trend of the 90\% confidence interval of $\theta_\tau$ in this table is the same as $\theta_\tau$ in table \ref{tb:OUBMCIRrej_u}, but the deviation is not as good as that.

\subsection{OUOUCIR Model}\label{sub:ououcir}
\begin{table}[H]
\caption{OUOUCIR model: Bias, Standard deviation and 90\% interval for parameters $\alpha_y, \alpha_x, \theta_x, \sigma_x, \alpha_\tau, \theta_\tau, \sigma_\tau, b_0, b_1, b_2$ with uniform prior use rejection approach.}
\begin{tabular}{ccrrrrccrrrr}
& $n$ & \tc{10} & \tc{20} & \tc{50} & \tc{100} & & $n$ & \tc{10} & \tc{20} & \tc{50} & \tc{100}\\
Par.& & & & & & Par.& & & & \\ \hline
\multirow{4}{*}{$\alpha_y$}
& bias & 0.005 & 0.002 & 0.002 & 0.003& \multirow{4}{*}{$\theta_\tau$}
& bias & 11.402 & 8.777 & 2.588 & 3.484 \\
& sd & 0.085 & 0.086 & 0.087 & 0.086& & sd & 12.649 & 12.101 & 11.678 & 14.117 \\
& 5\% & 0.019 & 0.016 & 0.013 & 0.018& & 5\% & 3.037 & 3.961 & 10.545 & 5.747 \\
& 95\% & 0.286 & 0.285 & 0.284 & 0.284& & 95\% & 44.453 & 44.068 & 49.130 & 52.252 \\ \hline
\multirow{4}{*}{$\alpha_x$}
& bias & 0.003 & 0.002 & 0.004 & 0.000& \multirow{4}{*}{$\sigma_\tau$}
& bias & 0.006 & 0.002 & 0.016 & 0.009 \\
& sd & 0.057 & 0.057 & 0.058 & 0.058& & sd & 0.289 & 0.287 & 0.285 & 0.290 \\
& 5\% & 0.010 & 0.013 & 0.010 & 0.010& & 5\% & 0.053 & 0.052 & 0.063 & 0.055 \\
& 95\% & 0.189 & 0.189 & 0.189 & 0.191& & 95\% & 0.951 & 0.950 & 0.948 & 0.956 \\ \hline
\multirow{4}{*}{$\theta_x$}
& bias & 0.014 & 0.003 & 0.059 & 0.001& \multirow{4}{*}{$b_0$}
& bias & 0.003 & 0.002 & 0.017 & 0.022 \\
& sd & 0.575 & 0.576 & 0.581 & 0.580& & sd & 0.575 & 0.584 & 0.585 & 0.571 \\
& 5\% & -0.912 & -0.906 & -0.922 & -0.895& & 5\% & -0.897 & -0.900 & -0.916 & -0.893 \\
& 95\% & 0.896 & 0.890 & 0.902 & 0.920& & 95\% & 0.898 & 0.906 & 0.907 & 0.891 \\ \hline
\multirow{4}{*}{$\sigma_x$}
& bias & 0.105 & 0.274 & 0.048 & 0.044& \multirow{4}{*}{$b_1$}
& bias & 0.008 & 0.002 & 0.007 & 0.015 \\
& sd & 0.390 & 0.356 & 0.327 & 0.359& & sd & 0.289 & 0.286 & 0.289 & 0.290 \\
& 5\% & 0.556 & 0.730 & 0.602 & 0.472& & 5\% & 0.055 & 0.055 & 0.052 & 0.042 \\
& 95\% & 1.840 & 1.875 & 1.673 & 1.642& & 95\% & 0.948 & 0.951 & 0.954 & 0.953 \\ \hline
\multirow{4}{*}{$\alpha_\tau$}
& bias & 0.016 & 0.016 & 0.010 & 0.002& \multirow{4}{*}{$b_2$}
& bias & 0.007 & 0.008 & 0.001 & 0.003 \\
& sd & 0.117 & 0.113 & 0.116 & 0.117& & sd & 0.288 & 0.288 & 0.287 & 0.286 \\
& 5\% & 0.024 & 0.024 & 0.021 & 0.018& & 5\% & 0.055 & 0.045 & 0.049 & 0.051 \\
& 95\% & 0.384 & 0.383 & 0.380 & 0.380& & 95\% & 0.951 & 0.947 & 0.950 & 0.947 \\ \hline
\end{tabular}
\label{tb:OUOUCIRrej_u}
\end{table}
In table \ref{tb:OUOUCIRrej_u}, the OUOUCIR model is more complex than the other three models, so the estimated results are not very well.
Only the $\alpha_\tau$ estimate much better in all parameters, because we want to our bias value and sd, will be smaller when size is bigger.

\begin{table}[H]
\caption{OUOUCIR model: Bias, Standard deviation and 90\% interval for parameters $\alpha_y, \alpha_x, \theta_x, \sigma_x, \alpha_\tau, \theta_\tau, \sigma_\tau, b_0, b_1, b_2$ with non-information prior and use rejection approach.}
\begin{tabular}{ccrrrrccrrrr}
& $n$ & \tc{10} & \tc{20} & \tc{50} & \tc{100} & & $n$ & \tc{10} & \tc{20} & \tc{50} & \tc{100}\\
Par.& & & & & & Par.& & & & \\ \hline
\multirow{4}{*}{$\alpha_y$}
& bias & 0.042 & 0.047 & 0.047 & 0.046& \multirow{4}{*}{$\theta_\tau$}
& bias & 0.450 & 0.699 & 1.698 & 0.924 \\
& sd & 0.146 & 0.153 & 0.152 & 0.153& & sd & 6.100 & 5.599 & 5.075 & 4.951 \\
& 5\% & 0.007 & 0.008 & 0.008 & 0.008& & 5\% & 21.029 & 21.455 & 20.773 & 21.617 \\
& 95\% & 0.451 & 0.455 & 0.458 & 0.450& & 95\% & 40.819 & 39.582 & 37.457 & 37.738 \\ \hline
\multirow{4}{*}{$\alpha_x$}
& bias & 0.031 & 0.032 & 0.033 & 0.027& \multirow{4}{*}{$\sigma_\tau$}
& bias & 0.147 & 0.178 & 0.137 & 0.178 \\
& sd & 0.101 & 0.103 & 0.099 & 0.100& & sd & 0.497 & 0.442 & 0.482 & 0.477 \\
& 5\% & 0.005 & 0.004 & 0.005 & 0.004& & 5\% & 0.026 & 0.026 & 0.032 & 0.022 \\
& 95\% & 0.296 & 0.304 & 0.301 & 0.297& & 95\% & 1.443 & 1.304 & 1.465 & 1.407 \\ \hline
\multirow{4}{*}{$\theta_x$}
& bias & 0.016 & 0.025 & 0.029 & 0.004& \multirow{4}{*}{$b_0$}
& bias & 0.026 & 0.014 & 0.006 & 0.016 \\
& sd & 1.001 & 0.999 & 1.015 & 1.023& & sd & 0.572 & 0.579 & 0.576 & 0.576 \\
& 5\% & -1.607 & -1.601 & -1.641 & -1.706& & 5\% & -0.905 & -0.908 & -0.910 & -0.890 \\
& 95\% & 1.653 & 1.674 & 1.714 & 1.666& & 95\% & 0.902 & 0.900 & 0.899 & 0.904 \\ \hline
\multirow{4}{*}{$\sigma_x$}
& bias & 0.615 & 0.384 & 0.285 & 0.307& \multirow{4}{*}{$b_1$}
& bias & 0.015 & 0.002 & 0.010 & 0.002 \\
& sd & 0.296 & 0.299 & 0.333 & 0.334& & sd & 0.287 & 0.287 & 0.290 & 0.288 \\
& 5\% & 0.109 & 0.284 & 0.318 & 0.305& & 5\% & 0.054 & 0.045 & 0.052 & 0.045 \\
& 95\% & 1.035 & 1.221 & 1.416 & 1.400& & 95\% & 0.950 & 0.947 & 0.954 & 0.946 \\ \hline
\multirow{4}{*}{$\alpha_\tau$}
& bias & 0.063 & 0.061 & 0.050 & 0.063& \multirow{4}{*}{$b_2$}
& bias & 0.002 & 0.001 & 0.004 & 0.023 \\
& sd & 0.196 & 0.195 & 0.201 & 0.190& & sd & 0.290 & 0.283 & 0.286 & 0.286 \\
& 5\% & 0.011 & 0.011 & 0.013 & 0.010& & 5\% & 0.050 & 0.054 & 0.053 & 0.049 \\
& 95\% & 0.596 & 0.585 & 0.613 & 0.565& & 95\% & 0.949 & 0.941 & 0.947 & 0.951 \\\hline
\end{tabular}
\label{tb:OUOUCIRrej_nu}
\end{table}

In table \ref{tb:OUOUCIRrej_nu}, although the deviation is not what we expected that keep getting smaller when the size gets larger, the range of the confidence interval is with our expectation.

\section{Empirical Data Analysis}\label{sec:emp}
Currently, we collect and analyze bat, fish, lizard, coral, foram and fig data from the literature.
We then fit our models into those data set and compare the fit of models.
We set prior parameters values $\alpha_y, \alpha_x, \alpha_\tau \sim$ exp(5), $\theta_x \sim$ N(0,1), $\tau \sim$ exp(3), $\sigma_x, \sigma_\tau \sim$ exp(2), $\theta_\tau \sim \chi^2_{30}$ and $b_0, b_1, b_2$ determine the uniform distribution range though the ordinary least squares (OLS) estimated value from the empirical data.
Under the ABC rejection approach, we run fifty thousand replicates and we set the tolerance rate 5\% for each model.

The overall result is shown in table \ref{tb:ranking_table}, the first column shows the trait we analyze while the last column shows the reference we use. The second, third, fourth and fifth column is the ranking of the models.
We collect data from the literature.
In the table \ref{tb:ranking_table}, the OUBMCIR, and OUOUCIR models are the best models or the second best model in our collect data.
\begin{table}[H]
\caption{The model selection by Bayes factor in Empirical Data}
\centering
\begin{tabular}{llllll}
 Data & $1^{st}$ & $2^{nd}$ & $3^{rd}$ & $4^{th}$ & References\\
  \hline
  bat & \textbf{oubmcir} & oubmbm & ououcir & ououbm &\cite{Aguirre02} \\
  lizard & \textbf{oubmcir} & \textbf{ououcir} & oubmbm & ououbm &\cite{Bonnie05}\\
  fish & oubmbm & \textbf{oubmcir} & ououbm & ououcir &\cite{crespi2002}\\
  lizard & \textbf{oubmcir} & oubmbm & ououcir & ououbm &\cite{molina2004}\\
  lizard & \textbf{oubmcir} & oubmbm & ououcir & ououbm &\cite{molina2004}\\
  lizard & \textbf{oubmcir} & oubmbm & ououcir & ououbm &\cite{niewiarowski04}\\
  coral & \textbf{oubmcir} & oubmbm & ououcir & ououbm &\cite{sanchez2003}\\
  foram & ououbm & \textbf{ououcir} & oubmcir & oubmbm &\cite{webster2002}\\
  fig & ououbm & oubmbm & ououcir & oubmcir &\cite{Weiblen04}\\\hline

\end{tabular}\label{tb:ranking_table}
\end{table}

For foram data in \cite{webster2002}, the best model is OUBMBM, the second best model is OUOUCIR, the third model is OUBMBM and the last model is OUBMCIR.
Their Bayes factors is shown in Table \ref{tb:BF_foram}.
From this table, we have the best model is OUOUBM because its Bayes factors are greater than one when comparing to other models.
Actually, the Bayes factor is 23.417 comparing to OUBMBM, is 20.960 comparing to OUBMCIR, and is 2.617 comparing to OUOUCIR.
The second best model is OUOUCIR because it has a Bayes factor of a value smaller than the best model (0.382 actually when comparing to OUOUBM) and has two Bayes factors greater than one (8.948 when comparing to OUBMBM and 8.009 when comparing to OUBMCIR).
Similarly, we observed that the OUBMBM as the third model and the last model is OUBMCIR.
\begin{table}[H]
\centering
\caption{Bayes factor table for foram dataset in \cite{webster2002}.}
\begin{tabular}{lrrrr}
  & OUBMBM & OUBMCIR & OUOUBM & OUOUCIR \\
   \hline
 OUBMBM & 1.000 & 0.895 & 0.043 & 0.112 \\
   OUBMCIR & 1.117 & 1.000 & 0.048 & 0.125 \\
   OUOUBM & 23.417 & 20.960 & 1.000 & 2.617 \\
   OUOUCIR & 8.948 & 8.009 & 0.382 & 1.000 \\\hline
\end{tabular}\label{tb:BF_foram}
\end{table}
We use the range of $K$ values proposed by \cite{kass1995} to compare the support between models for foram data in \cite{webster2002}. We see the third row in Table \ref{tb:BF_foram}, the values are 23.417, 20.960 and 2.617 that mean is the best model OUOUBM have stronger support than the OUBMBM, OUBMCIR, and OUOUCIR models.
When we see the second best model OUOUCIR that is to see the fourth row in Table \ref{tb:BF_foram}, it $K$ smaller than the best model the OUOUCIR is not better when comparing to OUOUBM, then $K$ is 8.948, when comparing to OUBMBM model, K between 1 and 3, K could not worth more than a bare mention for OUOUCIR by \cite{kass1995}.
Last, we compare OUOUCIR with OUBMCIR, the Bayes factor value, $K$, is 8.009, it explains the OUOUCIR have strong support than OUBMCIR in this data.
\begin{table}[H]
\centering
\caption{Bayes factor table for lizard dataset in \cite{Bonnie05}}
\begin{tabular}{lrrrr}
 & OUBMBM & OUBMCIR & OUOUBM & OUOUCIR \\
\hline
OUBMBM & 1.000 & 0.891 & 1.085 & 0.914 \\
OUBMCIR & 1.122 & 1.000 & 1.218 & 1.026 \\
OUOUBM & 0.921 & 0.821 & 1.000 & 0.842 \\
OUOUCIR & 1.094 & 0.975 & 1.187 & 1.000 \\ \hline
\end{tabular}\label{tb:BF_lizard_b05}
\end{table}
For fish data in \cite{Bonnie05}, the best model is OUBMCIR model, the second best model is OUOUCIR model, the third model is OUBMBM and the last model is OUOUBM model. The Bayes factor is shown in Table \ref{tb:BF_lizard_b05}.
From this table, we have the best model is OUBMCIR because its Bayes factors are greater than one when compare with other models.
In fact, the Bayes factor is 1.122 compared with OUBMBM, is 1.218 compared with OUOUBM and is 1.026 comparing to OUOUCIR. The second best model is OUOUCIR because it has a Bayes factor of the value smaller than the best model, is 0.975 when comparing to OUBMCIR, and has greater than other models, is 1.094 comparing to OUBMBM and is 1.187 comparing to OUOUBM.
In this data, every model is not significant for each other because of they Bayes factor of value, $K$, is between 1 and 3 that explain not worth more than a bare mention.
But the OUBMCIR and OUOUCIR models are the best top two in the lizard dataset in \cite{Bonnie05}.
This is we want to see a good result because we hope our new model is the best model for four models in the special data.
Although the best model can be selected from the Table \ref{tb:BF_foram} and Table \ref{tb:BF_lizard_b05}, it is not significant in the lizard data in \cite{Bonnie05}. Therefore, we analyzed the foram data in \cite{webster2002} because it has a clear difference for each model. That is, between two models have a model get more support in this data.

Next, we analyze coral data because new model OUBMCIR has a good result in the different methods.
Table \ref{tb:est} shows estimated values of various models under different methods.
Table \ref{tb:estimateb0} shows estimation of $b_0, b_1$ and $b_2$ by OLS, ABC-rejection and ABC-MCMC approach under this data and shows that 95\% confidence interval.
The estimated value of $b_0$, $b_1$ and $b_2$  are mean of every model posterior value, for different approach.
\begin{table}[H]
\centering
\caption{The estimator under coral data in \cite{Sanchez03}}
\begin{tabular}{clcccccccc}
 Method & \tc{Model} & $\alpha_y$ & $\sigma_x$ & $\tau$ & $\alpha_x$ & $\theta_x$ & $\alpha_\tau$ & $\theta_\tau$ & $\sigma_\tau$ \\
  \hline
\multirow{4}{*}{ABC-Rej}&OUBMBM & 0.200 & 1.113 & 0.336 & - & - & - & - & - \\
&OUOUBM & 0.415 & 1.125 & 0.333 & 0.210 & 0.227 & - & - & - \\
&OUBMCIR & 0.200 & 1.115 & - & - & - & 0.198 & 1.893 & 0.500 \\
&OUOUCIR & 0.211 & 1.089 & - & 0.203 & 0.162 & 0.209 & 2.259 & 0.502 \\
\end{tabular}
\label{tb:est}
\end{table}
\begin{table}[H]
\centering
\caption{The Beta estimator under coral data in \cite{Sanchez03}}
\begin{tabular}{ccccc}
  &  Model & \tc{$b_0$} & \tc{$b_1$} & \tc{$b_2$} \\
  \hline
OLS& $Y=X_1 + X_2$& -1.197 & 2.854 & 1.340 \\ \hline
\multirow{8}{*}{ABC-Rej}
  & \tc{oubmbm} & -1.182 & 3.227 & 1.920 \\
  & & \small(-2.951 , 0.550) & \small(0.275 , 5.584) &\small(-3.666 , 6.475) \\
  & \tc{ououbm} & -0.711 & 3.398 & 2.156 \\
  & &\small(-2.836 , 0.603)&\small(0.298 , 5.624)&\small(-3.788 , 6.555)\\
  & \tc{oubmcir} & -1.197 & 2.917 & 1.441 \\
  & &\small(-2.962 , 0.537)&\small(0.161 , 5.547)&\small(-3.717 , 6.427)\\
  & \tc{ououcir} & -1.177 & 2.829 & 1.387 \\
  &&\small(-2.944 , 0.568)&\small(0.163 5.529)&\small(-3.768 , 6.417)\\\hline
\end{tabular}
\label{tb:estimateb0}
\end{table}

\section{Conclusion}\label{sec:cls}
In this paper, we expand two models for the adaptive trait evolution and called them the OUBMCIR model and OUOUCIR model, respectively.
Due to the intractability of the likelihood function for the models, we make attempt to use Approximate Bayesian Computation to analyze data.
We propose relevant algorithm and derive the solution as explicitly as possible to simulate trait along the tree for each model.
Currently, our provide simulation to validate our model and analyze several empirical data sets with comparing the fit for the model using Bayes factors.
Currently, our results show that we have strong evidence to demonstrate the superiority of new models.
In table \ref{tb:ranking_table} we have nine datasets, the result seems to suggest that our new model could be a good and nice because as it provides a better fit than the existed models(OUBMBM and OUOUBM models) in empirical data.

And from the empirical data, we see the best model and second best model almost pointing to the new models OUBMCIR model and the OUOUCIR model.
Actually, the result is well but the method proposed by \cite{kass1995} makes the Bayes factor not significant in these data.
A part of future research that should be considered is using the others criterion of model selection, using the others prior distribution and collect the data to support our new models would be more useful.

\section*{References}

\bibliography{C012_2}

\end{document}